\documentstyle[11pt]{article}

\def\mb{\mbox}
\def\be{\beta}
\def\l{\lambda}
\def\a{\alpha}
\def\ap{\approx}
\def\t{\theta}
\def\p{\phi}
\def\g{\gamma}
\def\d{\delta}
\def\D{\Delta}
\def\L{\Lambda}
\def\s{\sigma}

\def\o{\over}
\def\le{\left}
\def\ri{\right}
\def\i{\hspace*{\parindent}}
\def\n{\noindent}
\def\r{\rightarrow}
\def\v{\vec}
\def\u{\underline}

\begin{document}

\title{The Behavior of a Spherical Hole in an Infinite Uniform Universe}

\author{Gilbert N. Lewis\\

Department of Mathematical Sciences\\
Michigan Technological University\\
Houghton, MI., 49931\\
{\tt lewis@math.mtu.edu}\\
and\\
Richard N. Lewis\\
deceased\\
coauthor of sections 1 and 2}
\date{}

\maketitle

\begin{abstract}

In this paper, the behavior of a spherical hole in an otherwise infinite and
uniform universe is investigated. First, the Newtonian theory is developed. The
concept of negative gravity, an outward gravitational force acting away from
the
center of the spherical hole, is presented, and the resulting expansion of the
hole is investigated. Then, the same result is derived using the techniques of
Einstein's theory of general relativity. The field equations are solved for an
infinite uniform universe and then for an infinite universe in which matter is
uniformly distributed except for a spherical hole. Negative pressure caused by
negative gravity is utilized. The physical significance of the cosmological
constant is explained, and a new physical concept, that of the gravitational
potential of a hole, is discussed. The relationship between the Newtonian
potential for a hole and the Schwarzschild solution of the field equations is
explored. Finally, the geodesic equations are considered. It is shown that
photons and particles are deflected away from the hole. An application of this
idea is pursued, in which a new cosmology based upon expanding holes in a
uniform universe is developed.  The microwave background radiation and Hubble's
Law, among others, are explained. Finally, current astronomical data are used
to
compute a remarkably accurate value of Hubble's constant, as well as estimates
of the average mass density of the universe and the cosmological constant.
\end{abstract}

\section{Introduction}
\label{intro}

\i In this paper, we consider a uniform, infinite universe with a spherical
hole. Ultimately, we intend to develop a new cosmological model for an
expanding
universe in which holes are the dominant feature. Geller and Huchra \cite[p.
900]{gandh} also indicate the importance of holes in the universe by writing
``...it probably makes more physical sense to regard the individual voids as
the
fundamental structures." Thus motivated, we consider the behavior of a hole in
an otherwise infinite and uniform universe.

We show first in section 2, using Newtonian theory, that there exists an
outward-directed gravitational force acting away from the center of the
spherical hole, and we develop a formula for this force. This force is referred
to as negative gravity. Then, in section 3, we solve the Newtonian differential
equations to discover the expansionary behavior of the hole and develop certain
formul{\ae} to be used later.

We show the same conclusion using the methods of Einstein's general relativity
in section 4. The derivation of the standard Schwarzschild solution for an
attracting point particle of mass $m$ is well known. The results are summarized
in section 4.1, and several consequences are mentioned which will be important
later when the techniques used in the solution of the Schwarzschild problem are
applied to a nonstatic distribution of mass. It is well known that particles
and
photons are attracted toward the mass. In the geodesic equations describing the
motion of particles and photons, if the mass were negative, we would conclude
that particles and photons would deflect away from the mass. Although negative
mass may not have any physical significance, it serves as an introduction to
the
main results of later sections. The concept of negative mass will be used in
terms of superimposing on a uniform universe this Schwarzschild solution for
negative mass in order to simulate a mass deficiency (hole) in an otherwise
uniform universe. Section 4.2 gives a new interpretation of the solution of
Einstein's field equations for a uniform distribution of mass. This solution
represents a flat geometry and reduces to the flat geometry of special
relativity for mass density decreasing to zero. Section 4.3 describes the
effect
of a spherical hole in a uniform universe. In these sections also, the physical
significance of the cosmological constant is determined.

If this were a linear universe, the two solutions from sections 4.2 and 4.3
could be superimposed. In a physically imaginable universe which is initially
uniformly filled with mass except for a spherical hole, particles and photons
are deflected away from the hole. This follows since the geometry of the
uniform
distribution is flat and does not affect the uniform motion of particles and
photons, while the negative Schwarzschild mass of section 4.2 would repel
matter
and photons. Loosely speaking, the hole repels matter and energy. Also in
section 4.3, a remarkably close connection is made between the Schwarzschild
solution and the newly discovered Newtonian potential for holes, which includes
the contribution of negative gravity. The purpose of section 4.4 is to show
this
using the fully nonlinear Einstein field equations. Finally, the geodesic
equations are considered, and it is shown that particles and photons are
deflected away from the hole. Section 4.5 summarizes the results.

Section 5 gives a development of a new cosmology, based upon expanding holes in
an infinite, uniform universe. This is based upon a paper by R. N. Lewis
\cite{rlewis}, in which it is shown how structure could form from the
interactions of nearby holes, resulting in the universe we observe today. The
microwave background radiation, caused by gravitational redshift, and Hubble's
Law are shown to be natural consequences of the model. We also discuss the
relative abundances of elements throughout the universe, Olbers' paradox, and
other observed phenomena.

Finally, in section 6, we use some of the ideas and equations previously
derived
in order to calculate Hubble's constant, the average mass density in the
universe, and the cosmological constant. Certain accurate astronomical data are
used in the calculations. The value derived for Hubble's constant is shown to
lie within the experimentally-determined range for that constant, providing
confidence that the underlying theory is correct.

\section{Negative gravity}

\i In this section, we consider a Newtonian approach to the gravitational force
field in an infinite universe filled with a uniform distribution of mass with
the exception of a spherical hole. Physicists usually consider the potential
and
the force acting on a particle anywhere in such an infinite universe to be
infinite, since a formal integration over space produces this result. However,
we apply symmetry in our calculations to produce cancellations, which leads to
a
finite force at any point. This application of symmetry is mathematically
equivalent to a Cauchy-type integration, without which the integral would be
infinite. Whether this approach is valid depends on its agreement with the
results of general relativity. Using results in later sections of this paper,
we
show this to be the case.

\subsection{Hypotheses}

\i We consider an infinite, uniform distribution of matter with mass from a
spherical region removed and evaluate the force field at any point. The three
main hypotheses we use are:
\begin{enumerate}
\item Gravity obeys the (Newtonian) inverse-square force law, and it acts over
arbitrarily large distances and time intervals.  The general relativistic
aspects of this problem are taken up later in this paper.
\item When necessary, matter will be considered to be distributed continuously
rather than discretely throughout space.  With this assumption, the
calculations
become tractable.  On the other hand, the difference in effect on a point mass
of a continuous distribution of matter versus a discretely distributed (atomic)
set of matter is negligible.  Indeed, the only essential aspect of the
distribution of matter which is used in the computations leading to the results
in section 2.3 is spherical symmetry.  Also, when convenient, we will assume
that the mass density is constant.
\item The universe is spatially infinite.  This, along with the constant mass
density of hypothesis (2), implies that the universe contains an infinite
amount
of mass.
\end{enumerate}

\subsection{Compilation of some results from classical Newtonian gravitational
theory}

\i Here, we present some results which depend only upon the inverse-square law
of Newtonian theory.  See Kittel, Knight, and Rudderman \cite{kittel} for
appropriate computations involving gravitational forces.  The following is a
list of the necessary facts and formul\ae :
\begin{enumerate}
\item The gravitational force due to a spherical shell of mass of radius $a$,
(infinitesimal) thickness $dr$, and mass density $\rho$ (mass per unit volume)
acting on a mass $m$ outside the shell ($r>a$) is the same as if all of the
mass
of the shell were concentrated at the center of the sphere. Mathematically,
this
is expressed as

$$\v{F} = -{{4\pi Ga^2m\rho dr}\o{r^2}}\v{e},$$

where $G$ is Newton's gravitational constant ($6.67$ x $10^{-11}$ m$^3$/kg
s$^2$), $r$ is the distance from the center of the sphere to the mass, and
$\v{e}$ is a unit vector from the center of the sphere toward the mass.  In the
case $r = a$, a slight modification of the calculations leading to the above
result yields $\v{F} = -2\pi Gm\rho dr\v{e}.$

\item If we have the same situation as in (1) above, but the mass $m$ is inside
the shell ($r < a$), then there is no gravitational force acting on the mass.
Therefore

$$\v{F}  = \v{0}.$$

\n This result generalizes to the case of a spherical shell (not
infinitesimally
thin) of inner radius $a_1$ and outer radius $a_2$.  The gravitational force
acting on $m$ inside $a_1$ is still $\v{0}$.  Furthermore, the above results
generalize by integration to a sphere of mass acting on a mass $m$.

\item The force due to a sphere of mass of radius $a$ and mass density $\rho$
acting on a mass $m$ outside the sphere ($r > a$) is the same as if all the
mass
were concentrated at the center.  Thus

$$\v{F} = -\left[\left({4\o 3}\right)\pi a^3\rho
\right]{{Gm}\o{r^2}}\v{e}.$$

\item If we have the same situation as in (3) but the mass $m$ is inside the
sphere ($r < a$), then all of the mass outside of $r$ contributes nothing to
the
force, and the mass inside of $r$ acts on $m$ as if it were concentrated at the
center.  Therefore

$$\v{F} = -\left[\left({4\o 3}\right)\pi r^3\rho\right]{{Gm}\o{r^2}}\v{e}
= -\left({4\o 3}\right)\pi r\rho Gm\v{e}.$$

\end{enumerate}

Finally, we mention that, for a finite, spherically symmetric (not necessarily
uniform) mass distribution, the gravitational force of it on a point mass does
not depend on the mass farther away from the center than the point mass, and is
the same as if all of the mass closer to the center than the point mass were
concentrated at the center.  We will see in the next section how this result
generalizes to infinite mass distributions, with unexpected results.

\subsection{Negative gravity}

\i The next question we ask is: what is the gravitational force on a point mass
$m$ situated at a positive distance $r$ from the origin due to a uniformly
distributed (infinite) mass external to a spherical hole with center at the
origin and radius $a$ ($r < a$)?  That is, what is the gravitational effect on
a test particle in a spherical hole in an otherwise uniform universe?  If we
consider the situation specified in the generalization of formula (2) in
section 2.2, and na\"{i}vely let $a_2 \r \infty$, then we would conclude that
the total force acting on $m$ is $\v{0}$.  However, this conclusion is not
accurate, since the total mass which lies outside the sphere of radius $a_2$ is
not spherically symmetric about $m$ (although it is spherically symmetric about
the center of the sphere), and does have an effect on $m$.

In order to analyze this case, consider the situation depicted in Figure 1.  We
take the origin of our coordinate system at $m$, with the center of the sphere
$S_2$, of radius $a$, at a point $r$ units to the right of $m$.  Then we draw a
sphere, $S_3$, with center at $m$ and radius $r + a$.  We also add a sphere,
$S_1$, concentric with $S_2$, and with radius $r$.  Since the mass external to
$S_3$ is spherically symmetric with respect to $m$, its total gravitational
force on $m$ is $\v{0}$.  Thus, the only effective force which will act on $m$
will be that due to the mass in the three-dimensional crescent-shaped region
(crescentoid), $R$, which lies inside $S_3$ and outside $S_2$ (see Fig. 1).

\eject
\vbox{\vspace{6in}}
\special{Figs1.eps}

In order to handle this situation, consider a totally uniform distribution of
matter (no hole) with a test mass $m$ at the origin.  By symmetry, there will
be
no force acting on $m$.  Also by symmetry, the net force on $m$ due to all mass
inside $S_3$ is $\v{0}$.  The force on $m$ due to all the mass in the spherical
shell between $S_1$ and $S_2$ is $\v{0}$, from the results of formula (2) and
its generalization in section 2.2.  Therefore, the mass in the region $R$ will
have a force on $m$ which will exactly counterbalance the force on $m$ due to
the mass inside $S_1$.  From section 2.2, formula (4), we find the force on $m$
due to the mass inside $S_1$ to be $-4\pi rGm\rho \v{e}/3$, where $\v{e}$ is a
unit vector from the center of $S_1$ toward $m$.  Thus, the force on $m$ due to
the mass in the shaded region (crescentoid) is

$$\v{F} = \left({4\o 3}\right)\pi rGm\rho \v{e}. \eqno (1)$$

\n This will also be equal to the force acting on $m$ in a uniform universe
with a hole (either $S_1$ or $S_2$), since the counterbalancing mass inside
$S_1$ is not present.

We note the interesting result that this force does not depend on $a$. Because
of this, in some applications, we can think of $a$ as being arbitrarily large,
and the crescentoid region in the figure as being arbitrarily far away from
$m$.  The matter (or absence of it) in the spherical shell between $S_1$ and
$S_2$ provides no contribution to the force on $m$. We can therefore say that
the force $\v{F}$ on $m$ given by the above formula is due to mass arbitrarily
far away from $m$. We also note that, as time passes, matter moves outward from
the hole, destroying the constancy of matter but preserving spherical symmetry
about the center of the hole.  The above comments hold in this case as well,
since the crescentoid region referred to above can be considered to be
arbitrarily far from the motion and unaffected by it. The formula for force
given above still applies, with $r$ a function of time.

At this point, we mention that a formal mathematical integration over all
space,
after applying symmetry considerations, yields the same result.

We note that the force can be written as

$$\v{F} = \left({{GmM}\o{r^2}}\right)\v{e},$$

\n where $M = (4/3)\pi r^3\rho$ is the mass deficiency of the hole (out to
$r$).
This shows that Newton's inverse-square force law holds for negative masses
(holes) acting on positive masses.

This formula gives the force $\v{F}$ at any point inside the hole ($r < a$).
However, spherical shells of matter of any radius greater than $r$ have no
gravitational effect on a mass $m$ at $r$, so we can think of $a$ as being
arbitrarily large. Thus, the above formula yields the outward force at any
point inside \u{or} outside the hole. For a point outside ($r > a$) the hole,
however, there will be an additional inward-directed force due to all mass
closer to the origin than $r$. This inward force is

$$\v{F}_2 = -{{[\left({4\o 3}\right)\pi (r^3 - a^3)\rho
]Gm}\o{r^2}}\v{e}. \eqno (2)$$

\n The sum of the above two forces yields the force due to a hole of radius $a$
in an otherwise infinite and uniform distribution of mass on a mass $m$ at
distance $r$ ($> a$) from the origin (that is, the force of a hole acting on a
mass outside the hole). This force is

$$\v{F} = {{\left({4\o 3}\right)\pi a^3\rho mG}\o{r^2}}\v{e} =
({{GmM_1}\o{r^2}})\v{e},$$

\n where $M_1 = (4/3)\pi a^3\rho$  is the mass deficiency of the hole.  This
again shows the applicability of Newton's inverse-square force law with one
negative and one positive mass.

It is tempting to say that Newton's inverse-square force law holds for two
negative masses as well.  However, we would need to define what we mean by a
force on a hole.  We simply mention that two holes would have an effect on each
other due to the above-mentioned force acting on the particles surrounding
the holes.  The net effect would be that the two holes would move toward each
other and become slightly distorted.  We conclude that positive masses attract
each other, as do negative masses (holes), but positive and negative masses
repel each other.

The implication of the above result is the following.  The infinite amount of
matter outside of a hole in an otherwise infinite and uniform distribution of
matter will gravitationally draw matter out of the hole.  We can loosely say
that the hole repels matter, just as a concentration of matter would attract
matter (section 2.2). We refer to this phenomenon as {\it negative gravity}. It
must be remembered that this term refers to the apparent repulsion of matter by
holes. In fact, the actual mechanism causing negative gravity is the positive
gravitational attraction of ''mass at infinity", that is, the mass in a
crescentoid region arbitrarily far from the hole. The term negative gravity
should \u{not} be taken to imply that the gravitational force between two
particles is repulsive; gravity is always attractive. As we shall see in
section
4, however, the energy-momentum tensor in the neighborhood of a hole can be
negative because of the above-described effect.

\subsection{Generalization}

\i An obvious generalization to make to the above calculations is the
following.
In a spherically symmetric (about the origin) universe, the net force acting on
a particle of mass $m$ at a distance $r$ from the origin is

$$\v{F} = \left({4\o 3}\right)\pi Gm\bar{\rho}\v{e},$$

\n where $\bar{\rho}  = {\rho}_1 - {\rho}_2$, is the difference in the average
densities of matter outside and inside $r$, respectively. The density
${\rho}_2$
is the mass divided by the volume of the region inside $r$. The density
${\rho}_1$ is the average mass density over the region external to the sphere
of
radius $r$.  We note that if ${\rho}_2 = 0$, then this formula reduces to the
formula in section 2.3.  If ${\rho}_1 = 0$, (i.e. if there is negligible mass
outside a sphere of large radius, or if the mass density decreases to 0 as $r$
increases), then the net force is inward, the force being given by formula (3)
in section 2.2.  Thus, the basic underlying idea for negative gravity to exist
is that $\bar{\rho} > 0$.  Furthermore, if $\bar{\rho} > 0$, we speculate that,
even in the absence of total spherical symmetry, there will be a net outward
force (from centers of holes) tending to pull things apart.

\section{A Newtonian approach to the behavior of a spherical hole in an
infinite
universe}

\i Now let us investigate the behavior of a spherical hole in an otherwise
infinite and uniform universe and form some conclusions about the rate of
expansion.  As we point out later, a slow expansion caused by negative gravity
does occur.  We model the problem mathematically and solve two specific
problems
(sections 3.1 and 3.2) which assume that the age of the universe under
consideration is infinite.  The conclusions show the relatively large time
intervals that are required for particles to move an appreciable distance.

Under the assumptions given above, the outwardly-directed negative
gravitational
force acts at each point of space. For a particle of mass $m$, the force was
given above by equation (1).  This formula holds no matter what $r$ is. In
other
words, it holds both inside ($r < a$) and outside ($r > a$) the hole, although
when the point is outside the hole, there is also an inward-directed force
$\v{F}_2$ [equ. (2)] due to all mass inside $r$. In fact, this formula holds
even in a nonuniform universe as long as the universe is spherically symmetric
about the origin and $\rho$ represents the average mass density.

\subsection{Universe of discrete atoms}

\i Let us suppose that we have a universe of discrete hydrogen atoms.  Such a
universe cannot be uniform (continuous) except in a statistical sense, but it
can be spherically symmetric about the origin.  Let us suppose that this is the
case and that an atom, $H_0$, lies at the origin. We also assume that the
average mass density, $\rho$, is one atom per cubic meter ($1.67$ x $10^{-27}$
kg/m$^3$). This is roughly on the same order of the value that current
observations of luminous matter lead us to believe. The presence of unobserved
matter would increase this value. The actual value of $\rho$  is immaterial to
the conclusion we wish to draw. Because of the density assumption, we assume
that the nearest other atoms are one meter (1 m) from the origin. To $H_0$, the
other atoms appear to be distributed spherically-symmetrically over all space
outside of a hole of radius one meter.

Now, let $H_0$ be displaced a distance $\a  << 1$ m in any direction. Let
$P_{\a}$ represent this configuration, with $P_0$ representing the original
spherically-symmetric universe. The hypotheses (a test particle in a hole in a
spherically-symmetric universe) of section 1 are satisfied. The outward force
given in that section acts on the atom $H_0$. Newton's Second Law of Motion
implies that

$$m{{d^2(r\v{e})}\o{dt^2}} = \left({4\o 3}\right) \pi rGm\rho \v{e},$$

\n where $r\v{e}$ is a position vector of the particle. The initial value
problem for $r$, therefore, is

$${{d^2r}\o{dt^2}} = \left({4\o 3}\right) \pi G\rho r, \eqno (3a)$$

$$r(0) = \a, \eqno (3b)$$

$${{dr}\o{dt}}(0) = 0. \eqno (3c)$$

\n Let $\beta = [(4/3)\pi G\rho ]^{1/2} = 6.84$ x $10^{-19}$ s$^{-1}$. Then the
solution of the initial value problem is

$$r(t) = \a cosh\beta t.$$

According to this differential equation model, the time, $t$, it takes for the
mass $m$ (hydrogen atom) to reach the edge of the hole ($r = 1$ m) and its
corresponding velocity depend on $\a$. For example, if $\a  = 0.1$ m, $t =
1.39$
x $10^{11}$ years. If $\a  = 0.001$ m, $t = 3.52$ x $10^{11}$ years.  In both
cases, since $v = dr/dt \approx  \beta r$, and since $r = 1$ m, then $v = 6.84$
x $10^{-19}$ m/s. Note that $t = \bigl(ln\{[1 + (1 - \a ^2)^{1/2}]/\a
\}\bigl)/\beta \approx [ln(2/\a )]/\beta \r  \infty$ as $\a  \r  0$. The
conclusion we arrive at is that it takes infinitely long to transform the
universe from configuration $P_0$ to configuration $P_\a$ for $\a << 1$ m.

There are two aspects of this model that need clarification.  The first is
that,
as the hydrogen atom $H_0$ moves, it also gravitationally affects all the
other atoms in the universe, displacing them slightly and destroying the
spherical symmetry of the universe outside the hole. To counter this problem,
we argue that the conclusion we reached at the end of the previous paragraph is
still valid. The reason to believe this is that the smaller the value of $\a$

is, the better the initial value problem $(3)$ models the physical situation.
In the next section, we present a model in which spherical symmetry is
preserved, so that this problem does not arise. The other remark to make is a
positive one. This model allows us to start with a uniform universe, $P_0$, at
time $t = -\infty$ (let $\a  \r  0$) instead of starting at a particular time,
$t_0$ (``Big Bang"), with a particular mass density and having to worry about
what happened before that time or about what caused things to happen.

\subsection{Uniform universe with a hole}

\i Now let us consider a uniform (rather than a discrete) distribution of mass
[$\rho (x,y,z) = 1.67$ x $10^{-27}$ kg/m$^3$] and suppose that, at time $t =
0$,
a spherical hole of radius $a$ is introduced, centered at the origin $O$.
$\{$Note: $a = [3/(4\pi )]^{1/3}$ m if we consider the problem of taking away
the equivalent mass of one atom at the origin.$\}$ Consider the force acting at
point $P$ where $r = r_1 \ge  a$.

\eject
\vbox{\vspace{5in}}
\special{Figs2.eps}

At time $t = 0$, we have the situation depicted in Figure 2. There is an
outward force (relative to $O$) acting on each particle (see below), so at a
later time we will have the situation depicted in Figure 3. At time $t = 0$,
the force acting on a particle at a distance $r = r_1$ from $O$ is

$$\v{F} = \v{F}_1 + \v{F}_2, $$

\n where [equ. (1)]

$$\v{F}_1 = \left({4\o 3}\right) \pi Gm\rho r_1\v{e}$$

\n is the outward force due to negative gravity of all mass outside $r = r_1$,
and [equ. (2)]

$$\v{F}_2 = -{{\left[\left({4\o 3}\right) \pi r_1^3 - \left({4\o
3}\right) \pi a^3\right]\rho Gm}\o{r_1^2}}\v{e}$$

\n is the inward force due to mass inside $r = r_1$, where $[(4/3)\pi r_1^3 -
(4/3)\pi a^3]\rho$ is the mass inside the sphere of radius $r_1$ centered at
$O$. At time $t = 0$,

$$\v{F}(0) = {{\left({4\o 3}\right) \pi a^3\rho Gm}\o{r_1^2}}\v{e},$$

\n which is outward-directed. The magnitude of $\v{F}(0)$ decreases with
increasing distance ($r_1$) from $O$.  At time $t = t_1 > 0$, the force acting
on a particle at distance $r$ from $O$ is

$$\v{F} = \v{F}_1 + \v{F}_2,$$

\n where

$$\v{F}_1 = \left({4\o 3}\right) \pi Gm\rho r\v{ e}$$

\n is the force due to negative gravity (it changes with $r$), and $\v{F}_2$ is
the inward force due to positive gravity of all matter closer to the origin
than $r$. Because of the initial outward-directed force, all particles will be
accelerated outward.  We will see later that this outward motion will continue
forever.

At this point, let us make a further assumption about the motion. We will
assume
that atoms do not pass each other on their outward journey as time
progresses. To justify this assumption, consider what would happen if one
particle were to pass another. The force $\v{F}_1$ on each would not change,
but the inward-directed force $\v{F}_2$ would.  The magnitude of $\v{F}_2$ on
the overtak\u{ing} particle would increase, due to the increased mass inside
the
sphere of radius $r$, thus decreasing the magnitude of the outward force
$\v{F}$. The magnitude of the force $\v{F}_2$ on the overtak\u{en} particle
would similarly decrease. The net effect of one particle passing another would
be a decrease in acceleration on the passing particle and an increase in
acceleration on the passed particle. Other particles farther away would see no
effect of this passing, so we just assume it does not occur. With this
assumption, we can write

$$\v{F}_2 = -{{\left[\left({4\o 3}\right) \pi r_1^3 - \left({4\o 3}\right)
\pi a^3\right]\rho Gm}\o{r^2}}\v{e}.$$

\n Because of the above assumption, the total amount of matter closer to $O$
than $r$ remains unchanged (the term in brackets in the numerator), but
$\v{F}_2$ has changed since $r$ in the denominator has increased from its
initial value $r_1$. Therefore

$$\v{F} = \v{F}_1 + \v{F}_2 = \left({4\o 3}\right) \pi Gm\rho \v{e}\left(r
+ {{a^3}\o{r^2}} - {{r_1^3}\o{r^2}}\right).$$

\n This is always an outward-directed force since the last parenthesized term
is
always positive. At time $t = 0$, it equals $a^3/r_1^2$ since $r = r_1$. When
$t
> 0$, then $r > r_1$, so the term in parentheses is greater than $a^3/r^2$.

Since position is given by $r\v{e}$, then Newton's Second Law gives

$$m{{d^2(r\v{e})}\o{dt^2}} = {{\left({4\o 3}\right) \pi Gm\rho \v{e}(r^3
-r_1^3 + a^3)}\o{r^2}},$$

\n or, with $\beta ^2 = (4/3)\pi G\rho$,

$${{d^2r}\o{dt^2}} = {{\beta ^2(r^3 - r_1^3 + a^3)}\o{r^2}}. \eqno (4)$$

\n But $v = dr/dt$, so

$${{d^2r}\o{dt^2}} = {{dv}\o{dt}} = {{dr}\o{dt}}\,{{dv}\o{dr}} =
v\,{{dv}\o{dr}}.$$

\n Therefore

$$v\, dv = {{\beta ^2(r^3 - r_1^3 + a^3)dr}\o{r^2}}.$$

\n The solution satisfying $v = 0$ and $r = r_1$ at $t = 0$ is

$$v = {{dr}\o{dt}} = \left\{ 2\beta ^2\left[{{r^2}\o 2} + {{r_1^3}\o r}
- {{a^3}\o r} + {{a^3}\o{r_1}} - \left({3\o
2}\right)r_1^2\right]\right\} ^{1/2},$$

\n or

$${{dr}\o{\left(r^2 + {{2r_1^3}\o r} - {{2a^3}\o r} + {{2a^3}\o{r_1}}
- 3r_1^2\right)^{1/2}}} = \beta\, dt. \eqno (5)$$

\n This is difficult to integrate except when $r_1 = a$ (a point at the edge of
the hole).  In that case,

$${{dr}\o{(r^2 - a^2)^{1/2}}} = \beta\, dt.$$

\n The solution satisfying $r(0) = a$, found by trigonometric substitution in
the integral of the above expression or by direct solution of the second order
differential equation [with $v(0) = 0$], is

$$r(t) = acosh\beta t,$$

$$v(t) = {{dr}\o{dt}} = a\beta sinh\beta t.$$

Note that, for $t >> 1, v = \beta r$ (the same formula was derived in section
3.1 for a discrete distribution of matter), which shows that Hubble's Law holds
in this (accelerating) expanding universe, with no need for introducing a
singularity (``Big Bang") at some finite time in the past. In addition, the
value of $\beta$ given here as the proportionality constant is surprisingly
close to the estimated value of the Hubble constant $H$. I propose that this
value for $\beta$ is the one that should be used for the Hubble constant, not
necessarily the numerical value which is dependent upon a more accurate
estimate
of the average density in the universe, but the formula given before
equation (4). See section 6 for a more accurate calculation of $\rho$ and hence
of $\beta$.

Now we consider equation (5) when $r_1 > a$. In equation (5), we let
$\gamma = 2(r_1^3 - a^3)/r_1^3$ and $r = r_1s$ and integrate, using $r(0) =
r_1$, to obtain

$$\beta t = \int_1^{r/r_1} {{s^{1/2}}\o{[(s - 1)(s^2 + s - \gamma
)]^{1/2}}}ds.$$

\n This can be written in terms of elliptic integrals of the first and third
kinds (see \cite[p. 265]{gandr}). We will not pursue this integral solution
further since we are able to derive the desired results directly from the
differential equation.

Now we consider the velocity of the particles. We have already shown that the
magnitude of $\v{F}(0)$ (and therefore of acceleration) is a decreasing
function of initial position. Thus, for a certain time interval after $t = 0$,
the speed of particles will be a decreasing function of position also. We will
assume that this situation holds for all $t > 0$.  Since particles that start
farther out always move slower, the particles which began closer to the origin
will begin to catch up to, but never overtake, the more distant particles.
Thus,
an outward moving spherical wave of matter starts to build up at the edge
of the hole.

Now let us find out how long the above process takes. Using values for the
constants given above, the time $t$ required for the spherical wave of matter
to
reach one light year in radius is given by

$$1 \mbox{light year} = 9.461 \mbox{x} 10^{15} \mbox{m} = r_1cosh\beta t
\approx
r_1{{e^{\beta t}}\o 2}.$$

\n If $r_1$ is the radius of a hole which holds the mass of one atom, then

$$r_1 = \left({3\o{4\pi}}\right) ^{1/3} \mbox{m} = .6204 \mbox{m},$$

\n and

$$t = \left({1\o{\beta}}\right) cosh^{-1}(1 \mbox{light year}/r_1) = 1.76
\mbox{x} 10^{12} \mbox{years}.$$

\n It will take $1.76$ trillion years to evacuate a hole of radius $1$ light
year. At that time, the velocity of the outwardly expanding wave would be $v =
a\beta sinh\beta t = 1.30$ x $10^{-2}$ m/s. Further calculations show that the
mass of a typical galaxy (our own, for instance) is about $1.97$ x $10^{30}$
kg/sun x $10^{11}$ suns/galaxy $= 1.97$ x $10^{41}$ kg. In the original
one-atom-per-cubic-meter universe, a sphere would need to have a radius of $r =
3.04$ x $10^{22}$ m $= 3.21$ x $10^6$ light years to contain that much mass.
Solving $r = r_1cosh\beta t$ for $t$ shows that it would take

$$t \approx \left({1\o{\beta}}\right)
\mbox{ln}\left({{2r}\o{r_1}}\right)  = 2.46 \mbox{x} 10^{12}\mbox{years}$$

\n in order to evacuate a hole of size large enough to have initially contained
the mass of a typical galaxy. The velocity of the outward-expanding wave at
that time would be $v = 4.78$ x $10^4$ m/s, still less than $0.02\%$ of the
speed of light. In analogy with section 3.1, we remark that the equation $r =
r_1cosh\beta t$ can be solved for $t$ as a function of $r$ and $r_1$. For a
given value of $r$, we see that $t \r  \infty$ as $r_1 \r 0$. This reconfirms
the conclusion we made in section 3.1 of the infinite age of the universe we
are
considering here. Later, we show how this model can be applied to develop a
cosmological model for our universe.

\section{A general relativistic approach to the behavior of a spherical hole in
an infinite universe}

\i Now we turn to an investigation of the problem of the behavior of a
spherical
hole in an infinite, uniform universe from the point of view of general
relativity. We will have occasion to use some of the previously-derived results
and will show that the Newtonian results are compatible with general
relativity.

\subsection{Schwarzschild solution; conclusion for negative mass}

\i The standard line element of Riemannian geometry is given as $ds^2 = g_{\a
\be}dx^{\a}dx^{\be}$, where the Einstein summation convention is used, and $\a$
and $\be$ take on values 1 through 4. The $g_{\a \be}$ are the elements of the
metric tensor. Because the problem we are considering exhibits spherical
symmetry, the coordinates $x^{\a}$ will be identified with the standard
spherical coordinates of special relativity:

$$x^1 = r, \  x^2 = \t , \  x^3 = \p , \  \mb{and} \ x^4 = ct,$$

\n where $c$ is the speed of light in a vacuum. Depending on context, the
origin
will be taken to be either the center of an attracting mass or, when dealing
with a hole in a uniform universe, the center of the hole. A standard solution
of Einstein's equations, which is now well known, was first found by
Schwarzschild in 1916. For a particle of mass $m$ in an otherwise empty space,
the static line element can be reduced to \cite[sec. 94-95]{tol}

$$ds^2 = -e^{\l}dr^2 - r^2d\t ^2 - r^2 sin^2\t d\p ^2 + e^{\nu}c^2dt^2, \eqno
(6)$$

\n where $\l$ and $\nu$ are functions of $r$ only. The relativistic field
equations are \cite[equ. (9.79)]{abs}

$$R^{\a \g} - Rg^{\a \g}/2 + \L g^{\a \g} = CT^{\a \g}. \eqno (7)$$

\n $R^{\a \g}$ is the contracted Riemann-Christoffel tensor, $R$ is a scalar
contracted from $R^{\a \g}$, $T^{\a \g}$ is the energy-momentum tensor, $\L$ is
the cosmological constant introduced by Einstein, and $C$ is a constant which
is related to the constant of gravitation in Newton's theory. $C$ is usually
evaluated when Poisson's equation for the potential is used as a first
approximation. This is not done here for two reasons: (1) a different potential
[that for holes; see equ. (8b)] is used, and (2) the value of $C$ is later
derived independently.

Later, the gravitational force acting on a particle of mass $m$ in a uniform
universe with a spherical hole whose center is at the origin will be
considered.
The results of the Newtonian theory are presented here for purposes of
comparison with the results of general relativity. Earlier, we found the
outward
Newtonian force at any point to be [see equs. (1) and (2)]

$$\v{F}(r) = \le\{
     \begin{array}{cl}
     {{4\pi Gm\rho _0r}\o 3}\v{e},     &\mb{inside the hole},\\ &\\
     {{4\pi Gm\rho _0\le(r - {{r_1^3 - a^3}\o{r^2}}\ri)}\o 3}\v{e},
&\mb{outside the hole}.
     \end{array}
     \ri. \eqno (8a)$$

\n This represents a conservative force field with corresponding potential

$$\phi (r) = \le\{
     \begin{array}{cl}
     -{{4\pi G\rho _0r^2}\o 6},     &\mb{inside the hole},\\ &\\
     -{{4\pi G\rho _0\le({{r^2}\o 2} + {{r_1^3 - a^3}\o r}\ri)}\o 3},
&\mb{outside the hole}.
     \end{array}
     \ri. \eqno (8b)$$

\n While this may not be the standard way of defining $\phi (r)$, the ultimate
use to which this is put is as a guide to solve the field equations. The
conclusion of that venture in no way depends upon this Newtonian potential.
Moreover, the similarity below between this Newtonian potential for holes and
the Schwarzschild solution is certainly more than coincidental. From equation
(8b), we finally note that, either inside or outside the hole, direct
differentiation yields

$$\nabla ^2\phi  = -4\pi G\rho _0.$$

\n This is the negative of the value found for the potential of an attracting
mass \cite[equ. (9.3)]{abs}.

Now consider an isolated mass at the origin. The equations (7) become
\cite[sec. 82]{tol}

$$CT{^1}{_1} = e^{-\l}\left({{\nu '}\o r} + {1\o{r^2}}\right) -
{1\o{r^2}} + \L , \eqno (9a)$$

$$CT{^2}{_2} = CT{^3}{_3} = e^{-\l}\left[{{\nu ''}\o 2} - {{\l '\nu '}\o 4}
+ {{\nu '^2}\o 4} + {{\nu ' - \l '}\o{2r}}\right] + \L, \eqno (9b)$$

$$CT{^4}{_4} = -e^{-\l}\left({{\l '}\o r} - {1\o{r^2}}\right) -
{1\o{r^2}} + \L, \eqno (9c)$$

\n where $T{^\a}{_\a}$ are the only possible nonzero elements of the
energy-momentum tensor. Only perfect fluids of density $\rho $ and pressure $p$
will be considered, so that the energy-momentum tensor reduces to

$$T{^\mu}{_\be} = \left(\rho  +
{p\o{c^2}}\right)\left({{dx^{\mu}}\o{ds}}\right)\left({{dx^{\a}}\o{ds}}\right)g_
{\a
\be} - \left({p\o{c^2}}\right)\d {^\mu}{_\be}. \eqno (10)$$

As will be shown later, the cosmological constant, $\L$, enters into the
equations through the influence of distant matter. In the Schwarzschild model,
special relativity boundary conditions at infinity are usually applied to
obtain
the value zero for $\L$. A nonzero value for $\L$ is included in these
calculations only for consideration in later models where distant matter cannot
be ignored. In the empty space surrounding the mass, the pressure and density
are zero, and so $T{^\a}{_\a} = 0$. At this point, standard calculations on
equations (9) yield the solution for the metric

$$e^{-\l} = e^{\nu} = 1 - {{2m}\o r} - {{\L r^2}\o  3}, \eqno (11)$$

\n where $m$ is the mass of the attracting object. Thus, the Schwarzschild line
element (6) becomes

$$ds^2 = -{{dr^2}\o{(1 - {{2m}\o r} - {{\L r^2}\o 3})}} - r^2d\t ^2 -
r^2sin^2\t d\phi ^2 + c^2(1 - {{2m}\o r} - {{\L r^2}\o 3})dt^2. \eqno (12)$$

Following Tolman \cite[sec 83]{tol}, we can write the geodesic equations for
a moving particle as

$${{d^2x^{\s}}\o{ds^2}} + \{ \mu \be ,\s \}
\left({{dx^{\mu}}\o{ds}}\right)\left({{dx^{\be}}\o{ds}}\right) = 0, \eqno
(13)$$

\n where the term in braces is the Christoffel symbol of the second kind.
Again, standard calculations yield the geodesics for a particle or for a
photon.
For a particle, we find that ($h$ is a constant, $h = r^2d\phi /ds$)

$${{d^2r}\o{ds^2}} = -{m\o{r^2}} - {{3mh^2}\o{r^4}} +
{{h^2}\o{r^3}} + {{\L r}\o 3}, \eqno (14a)$$

\n or, for a radially-traveling particle,

$${{d^2r}\o{ds^2}} = -{m\o{r^2}} + {{\L r}\o 3}. \eqno (14b)$$

\n The acceleration in (14b) is negative for positive $m$ and $\L  = 0$. Thus,
a particle is accelerated inward for an attracting body.

For a photon, we can also calculate the geodesics. Standard asymptotic analysis
shows that the photon is deflected as it passes the attracting mass. To first
order in $m/R$, where $R$ is the distance of closest approach to the origin,
the
angle $\phi$ satisfies

$$\phi  = -{{\pi}\o 2} - {{2m}\o R} \eqno (15a)$$

\n for an incoming photon, and

$$\phi  = {{\pi}\o 2} + {{2m}\o R} \eqno (15b)$$

\n for an outgoing photon. The photon has been deflected from the straight line
path a total of

$$\D \phi  = {{4m}\o R} \eqno (15c)$$

\n toward the source $m$, which agrees well with observations made during solar
eclipses.

The conclusion to be drawn from equations (14b) and (15c) is that, for
negative $m$ and $\L \ge  0$, particles are accelerated away from the mass
($d^2r/ds^2 > 0$, see \cite[pp. 337-338]{abs}), and photons are deflected away
from the mass ($\D \phi  < 0$). Particles and photons are deflected away from
the origin due to an outward directed Newtonian force or due to an
outward-curving geometry. In other words, mass deficiency repels matter and
energy.

\subsection{Geometry of a homogeneous universe}

\i Consider a spatially infinite universe with matter uniformly distributed
with
density $\rho _0$. By spherical symmetry and homogeneity, the line element has
the form [equ. (6)]

$$ds^2 = -e^{\l}dr^2 - r^2d\phi ^2 - r^2sin^2\t d\phi ^2 + e^{\nu}c^2dt^2,
\eqno (16)$$

\n where $\l$  and $\nu$ are constants independent of $r$ and $t$. Einstein's
field equations for a perfect fluid at constant pressure $p_0 $ and constant
density $\rho _0$ reduce to [see equs.(9) and (10)]

$$C{{p_0}\o{c^2}} = -{{e^{-\l}}\o{r^2}} + {1\o{r^2}} - \L , \eqno
(17a)$$

$$C{{p_0}\o{c^2}} = -\L , \eqno (17b)$$

$$C\rho _0 = {{e^{-\l}}\o{r^2}} - {1\o{r^2}} +\L , \eqno (17c)$$

\n so that (see \cite[equ. after (4.206)]{mcv})

$$e^{-\l} = 1,\ \l  = 0, \eqno (18a)$$

$$C{{p_0}\o{c^2}} = -\L , \eqno (18b)$$

$$C\rho _0 = \L , \eqno (18c)$$

$${{p_0}\o{c^2}} + \rho _0 = 0. \eqno (18d)$$

\n Equation (18c) shows that the cosmological constant $\L$ is nonzero. This
shows the importance of the boundary conditions at infinity. A zero value for
the cosmological constant is derived using special relativity (no mass)
boundary
conditions at infinity. A nonzero value is found by assuming uniformly
distributed mass at infinity. $\L$ must be positive in order that $\rho _0$ be
positive. The pressure $p_0$ will then be negative. This is extremely
counterintuitive. However, as we pointed out earlier, the introduction of a
hole
into an infinite, homogeneous, Newtonian universe causes matter to move away
from the hole. This explains the negative value of the pressure, since pressure
measures the tendency of matter to fill a hole. This concept represents a
significant change in cosmological thought. An infinite, uniform, static
universe is now allowed, along with negative pressure, although such a universe
is unstable (see secs. 4.4 and 5). See  \cite[p. 361]{abs}, for example, for a
contrasting view. Hawking \cite[p. 3]{haw} states the contrasting case even
more
strongly: ``We now know that the universal attractive nature of gravity is
inconsistent with a static infinite universe."

All three of $p_0$, $\rho _0$, and $\L$  are extremely small in magnitude.
Later, the value of $C = 4\pi G/c^2$ is found. Using the values $G = 6.67$ x
$10^{-11}$ m$^3/$kg\, s$^2, \rho _0 = 1.67$ x $10^{-27}$ kg/m$^3$, and $c =
3.00$ x $10^8$ m/s, then $\L = 1.56$ x $10^{-53}$ m$^{-2}$. The value of
density
is an estimate based upon observations of luminous matter and should provide a
lower bound for $\L$. The presence of even vast amounts of unobserved matter
would increase its value but not change the conclusions drawn. See section 6
for
a more accurate determination of $\L$.

In order to determine $\nu$ from (16), consider a radially-traveling photon
($ds = d\phi  = d\t  = 0$) such that

$${{dr}\o{dt}} = ce^{\nu /2} = {c\o i} \eqno (19)$$

\n represents the speed of light in a medium of density $\rho _0$, and $i$ is
the index of refraction. $e^{\nu /2}$ is a reduction factor for the speed of
light traveling through the medium, and $e^{\nu /2} < 1$. For a fixed medium,
this will be constant. Thus the speed of light in the medium is $ce^{\nu /2}$.
As $\rho _0$ decreases to zero, $e^{\nu /2}$ increases to one. For the low
density problem we envision here, $e^{\nu /2} \approx 1$. Therefore, we will
assume $e^{\nu /2} = 1$ and not concern ourselves with the difference.

It should also be recognized from the above equations that the cosmological
constant $\L$  is proportional to the density. This interpretation will provide
a justification for the appearance of that constant in the field equations for
cosmological purposes. It also gives a cosmological meaning to that constant
and will show the effect of distant matter on the behavior of particles.
Finally, in a sparse universe of discrete atoms, there will be no kinetic
interactions between atoms, and it is permissible to take $p_0 = 0$, with $\L =
4\pi G\rho_0/c^2$.

\subsection{Geometry of a uniform universe with a spherical hole; motion of
particles and photons inside the hole}

\i Now consider the same homogeneous universe with a spherical hole of radius
$a$. Take the origin at the center of the hole. The line element will now have
the form \cite[sec. 94]{tol}

$$ds^2 = -e^{\l}dr^2 - r^2d\t ^2 - r^2sin^2\t d\phi ^2 + e^{\nu}c^2dt^2, \eqno
(20)$$

\n where $\l$ and $\nu$  are functions of $r$ and $t$. The universe is no
longer static, but it can still be considered to be spherically symmetric.
Physically, think of arriving at this state by superimposing on a homogeneous
universe a negative mass $m$ spread out over a sphere of radius $a$, or,
equivalently, deleting a sphere of radius $a$ and mass $m$ from the uniform
universe. Mathematically, think of linearizing the field equations and
superimposing the Schwarzschild exterior solution with negative mass, $m < 0,\
e^{-\l} = 1 - 2m/r - \L r^2/3$, onto the flat geometry of a uniform universe
given in section 4.2, $e^{\l} = 1$. The field equations are (see \cite[sec.
98]{tol}; compare with equ. (9); note that dots and primes represent partial
derivatives with respect to $t$ and $r$, respectively)

$$C\left[-\left(\rho  +
{p\o{c^2}}\right)e^{\nu}c^2\left({{dt}\o{ds}}\right)^2 +
{p\o{c^2}}\right] = e^{-\l}\left({{\l '}\o r} - {1\o{r^2}}\right) +
{1\o{r^2}} - \L , \eqno (21a)$$

$$C\left[\left(\rho  +
{p\o{c^2}}\right)e^{\l}\left({{dr}\o{ds}}\right)^2 +
{p\o{c^2}}\right] = -e^{-\l}\left({{\nu '}\o r} + {1\o{r^2}}\right) +
{1\o{r^2}} - \L , \eqno (21b)$$

$$ C\left[\left(\rho  + {p\o{c^2}}\right)r^2\left({{d\t}\o{ds}}\right)^2 +
{p\o{c^2}}\right] = C\left[\left(\rho  + {p\o{c^2}}\right)r^2sin^2\t
\left({{d\phi}\o{ds}}\right)^2 + {p\o{c^2}}\right] = $$ $$-e^{-\l}\left[{{\nu
''}\o 2} - {{\l '\nu '}\o 4} + {{\nu '^2}\o 4} + {{\nu ' - \l '}\o{2r}}\right]
+
{{e^{-\nu}}\o{c^2}}\left({{\ddot{\l}}\o 2} + {{\dot{\l}^2}\o 4} - {{\dot{\l}
\dot{\nu}}\o 4}\right) - \L, \eqno (21c)$$

$$C\left[\left(\rho  +
{p\o{c^2}}\right)e^{\nu}\left({{dr}\o{ds}}\right)\left(c{{dt}\o{ds}}\right)
\right] = e^{-\l}{{\dot{\l}}\o {rc}}, \eqno (21d)$$

\n where $\rho$ and $p$ represent density and pressure. These equations can be
solved exactly in the interior of the hole where the energy-momentum tensor is
zero. Equations (21a) and (21b) imply $\l ' = -\nu '$. Choose $\l  = -\nu $.
Also, equation (21d) implies $\dot{\l}  = 0$. Birkhoff's Theorem \cite[sec.
99]{tol} then requires

$$e^{\nu} = e^{-\l} = 1 - {{2m}\o r} - {{\L r^2}\o 3}, \eqno (22a)$$

\n since the equations (21) reduce to the same form as equations (9) [see equ.
(11)]. Regularity at $r = 0$ requires $m = 0$. Therefore,

$$e^{\nu} = e^{-\l} = 1 - {{\L r^2}\o 3}. \eqno (22b)$$

In light of equation (15c), with $m = 0$, the $\L$  term has no influence on
photons which pass through the hole in a straight Euclidean line. However, with
$m = 0$ and $\L > 0$, equation (14b) implies that particles are accelerated
outward. By continuity, this outward acceleration must extend at least part way
into the matter-filled region outside the hole. We wonder whether the
singularity at $r = \sqrt{3/\L}$ in equation (22b) provides an upper limit to
the expansion of holes in a uniform universe.

In the Newtonian theory, inside the hole, we have [equs. (8)] $\phi = -4\pi
G\rho _0r^2/6, \v{F} = -m\v{\nabla}\phi = (4/3)\pi Gm\rho _0r\v{e}$, and

$$e^{\nu} \ap  1 + {{2\phi}\o{c^2}} = 1 - \left({{4\pi G\rho
_0}\o{c^2}}\right){{r^2} \o 3} = 1 - {{\L r^2}\o 3} \eqno (23a)$$

\n [from equ. (22b)], indicating that $\L = 4\pi G\rho _0/c^2$, and $C = 4\pi
G/c^2$. This is $-1/2$ times the value usually used for C, which is derived for
the potential of an attracting point mass. The equation $e^{\nu} \ap  1 + 2\phi
/c^2$ is derived in \cite[sec. 4.3]{abs} or \cite[sec. 80]{tol} for a time
independent metric, which exists in the interior of the hole. See \cite[p.
338]{abs} where it is suggested how to calculate $\L$ experimentally, thus
yielding a reliable value for $\rho _0$.

Before trying to solve the field equations (21) outside of the hole, where
matter is in motion and the metric is not static, let us make a connection
between this and the Newtonian theory. There is the approximation given above

$$e^{\nu} \ap  1 + {{2\phi}\o{c^2}}, \eqno (23b)$$

\n where $\phi$ is the Newtonian potential. We assume that this is
approximately
valid in the region exterior to the hole even though the metric is not static
there. The force acting on a particle of mass $m$ at any point outside the hole
is [see equ. (4a)]

$$\v{F} = \left({4\o 3}\right)\pi Gm\rho _0\v{e}\left[r - {{r_1^3 -
a^3}\o{r^2}}\right]. \eqno (24a)$$

\n Since the bracketed term is positive ($r \geq  r_1 \geq  a$), the force is
outward-directed. Thus, as time increases, the hole expands outward and
individual particles move away from the origin. The corresponding potential is

$$\phi  = -\left({4\o 3}\right)\pi G\rho _0\left[{{r^2}\o 2} + {{r_1^3 -
a^3}\o r}\right]. \eqno (24b)$$

\n Note that $\phi$  decreases toward $-\infty$ as $r$ increases
($\v{\nabla}\phi$ points in the same direction as $-\v{e}$), so that large $r$
values correspond to large negative potentials. Using this value for $\phi$, we
see that $e^{\nu}$ from equation (23) should be of the form

$$e^{\nu} \ap  1 + {{2\left[-\left({4\o 3}\right)\pi G\rho
_0\left({{r^2}\o 2} + {{r_1^3}\o r} - {{a^3}\o
r}\right)\right]}\o{c^2}} \eqno (25a)$$

$$= 1 - \left({{4\pi G\rho _0}\o{c^2}}\right){{r^2}\o 3} +
2\left[\left({4\o 3}\right)\pi \rho _0a^3\right]\left({G\o{c^2r}}\right) -
2\left[\left({4\o 3}\right)\pi \rho _0r_1^3\right]\left({G\o{c^2r}}\right).
\eqno (25b)$$

\n Earlier, we showed that, for a particle which started at
$r_1 = a$, the position in the Newtonian theory at any later time $t$ is $r =
acosh\be t,\ \be ^2 = 4\pi G\rho _0/3$. In addition, if $r_1 > a$ then $r >
acosh\be t$, but $r/acosh\be t \r 1$ as $t \r \infty$. It is reasonable to
conclude that $r_1 = f_1(r,t)r$ where $0 < f_1 \leq 1$, so that the last term
in
equation (25b) can be replaced with a general function $\g (r,t)$ which would
be
expected to lie asymptotically between $1/r$ and $r^2$ as $r \r \infty$. We
identify in the first coefficient of $1/r$ in equation (25b) the term $4\pi
a^3\rho _0/3$ which is the mass $M$ which was initially removed from an
infinite
homogeneous universe of constant density $\rho _0$ to make a hole of radius $a$
centered at the origin. Letting $m = -GM/c^2$ \cite[equs. (8.28), (8.33)]{abs}
or \cite[equ. (5.118)]{mcv}, this term is exactly $-2m/r$ which corresponds to
the term in the Schwarzschild exterior solution with negative mass [see equ.
(11)]. The coefficient of $r^2/3$ is exactly the $\L$ value calculated above.
Thus,

$$e^{\nu} = 1 - {{\L r^2}\o 3} - {{2m}\o r} + \g (r,t), \eqno (25c)$$

\n which differs from the Schwarzschild solution only in the $\g$ term. It
should also be noticed that the Schwarzschild solution, except for the $\g$
term, corresponds exactly with the Newtonian potential for holes. In
particular,
the $-2m/r$ term corresponds to the mass which was initially removed to form
the
hole of radius $a$ (negative Schwarzschild mass), and the $-\L r^2/3$ term
corresponds to the effect of negative gravity, that is, to mass at infinity.

The geodesic equations (13) now become more complicated, so the analysis of the
motion of photons and particles is made more complex. Thus, at this point, only
an educated guess as to the motion of photons and particles outside the hole
can
be made. Because of the outward Newtonian force, we claim that particles move
outward in the relativistic theory as well. This conclusion also follows from
equation (14b) if we neglect the $\g$ term. In fact, as was remarked earlier,
this conclusion follows from continuity at least near the edge of the hole.
Also, in comparison with equation (15c), we claim that the trajectories of
photons are bent outward in the matter-filled region exterior to the hole
because of the negative $m$ [negative Schwarzschild mass---see discussion after
equ. (15c)] referred to after equation (25b) (again neglecting $\g$). Thus, the
linear theory (superposition of solutions from section 4.2 and from this
section, neglecting $\g$) gives the conclusion that a hole repels matter and
energy. In the Newtonian theory, we showed that the outward acceleration was
due
to the effect of ``mass at infinity", that is, only that mass outside of a
sphere of arbitrarily large radius. From equation (14b) with $m = 0$, it can be
seen that the outward acceleration acting on particles is due to the $\L$ term.
Thus the cosmological constant $\L$ is associated with the distant mass, which
shows that Mach's principle \cite[p. 339]{abs}, at least insofar as it deals
with radial forces, is apparently incorporated into Einstein's theory through
the cosmological constant. This conclusion contrasts with conventional thinking
\cite[p. 371]{abs}.

\subsection{Solution of Einstein's geodesic equations in the matter-filled
universe exterior to the hole}

\i Now, consider the equations (21) outside of the hole, where matter is
present. First, it seems likely that zero is a lower bound for the sum $\rho  +
p/c^2$ [see equ. (18d)]. As matter moves away from the hole, the hole (where
$\rho  + p/c^2 = 0$) will expand. In addition, as matter builds up at the edge
of the hole and beyond, $\rho$ will increase there, but $p$ will also increase
due to the increased kinetic activity of the matter there. We will take this to
be an additional requirement:

$$\rho  + p/c^2 \ge  0. \eqno (18d')$$

Next, notice that subtraction of equation (21b) from equation (21a) results in

$$-C\le(\rho + {p\o{c^2}}\ri)\le[e^{\nu}c^2\le({{dt}\o{ds}}\ri)^2 +
e^{\l}\le({{dr}\o{ds}}\ri)^2\ri] = e^{-\l}\le({{\l ' + \nu '}\o r}\ri). \eqno
(26)$$

\n Since $C > 0$ (sec. 4.3) and $\rho + p/c^2 > 0$, then $\l ' + \nu ' < 0$.
Compare with \cite[equ. (9.121)]{abs}. Let us require further that $\l ' < 0$
and $\nu ' < 0$. Also note from equation (21d) that $\dot{\l}(dr/ds)(dt/ds) >
0$.

Now the geodesic equations (13) will be considered. These reduce to (for $\s =
1,2,3,4$)

$${{d^2r}\o{ds^2}} + \dot{\l}\le({{dr}\o{ds}}\ri)\le({{dt}\o{ds}}\ri) + {{\l
'}\o 2}\le({{dr}\o{ds}}\ri)^2 - rsin^2\t e^{-\l}\le({{d\p}\o{ds}}\ri)^2 - $$

$$re^{-\l}\le({{d\t}\o{ds}}\ri)^2 + \le({{e^{\nu -\l}\nu 'c^2}\o 2}\ri)
\le({{dt}\o{ds}}\ri)^2 = 0,  \eqno (27a)$$
\smallskip

$${{d^2\t}\o{ds^2}} + {2\o r}\le({{dr}\o{ds}}\ri)\le({{d\t}\o{ds}}\ri) -
sin\t cos\t \le({{d\p}\o{ds}}\ri)^2 = 0, \eqno (27b)$$
\smallskip

$${{d^2\p}\o{ds^2}} + {2\o r}\le({{dr}\o{ds}}\ri)\le({{d\p}\o{ds}}\ri) +
2cot\t \le({{d\p}\o{ds}}\ri)\le({{d\t}\o{ds}}\ri) = 0, \eqno (27c)$$
\smallskip

$$c{{d^2t}\o{ds^2}}+ \nu 'c\le({{dr}\o{ds}}\ri)\le({{dt}\o{ds}}\ri) +
\le({{e^{\l -\nu}\dot{\l}}\o{2c}}\ri)\le({{dr}\o{ds}}\ri)^2  +
\le({{\dot{\nu}c}\o 2}\ri)\le({{dt}\o{ds}}\ri)^2 = 0. \eqno (27d)$$

\n One solution of equation (27b) is $\t  = \pi /2$ (that is, motion takes
place
entirely in that plane), and the corresponding solution of equation (27c) is
$d\p /ds = h/r^2$, where $h$ is a constant. Using these, equations (27a) and
(27d) are simplified to

$${{d^2r}\o{ds^2}} + \dot{\l}\le({{dr}\o{ds}}\ri)\le({{dt}\o{ds}}\ri) +
{{e^{-\l}}\o r}\le[{{\l 'r}\o 2}e^{\l}\le({{dr}\o{ds}}\ri)^2 - {{h^2}\o{r^2}} +
{{\nu 'r}\o 2}e^{\nu}c^2\le({{dt}\o{ds}}\ri)^2\ri] = 0,
\eqno (28a)$$
\smallskip

$${{d^2t}\o{ds^2}} + \nu '\le({{dr}\o{ds}}\ri)\le({{dt}\o{ds}}\ri) +
{{e^{-\nu}}\o{2c^2}}\le[\dot\l e^{\l}\le({{dr}\o{ds}}\ri)^2 + \dot\nu
e^{\nu}c^2\le({{dt}\o{ds}}\ri)^2\ri] = 0. \eqno (28b)$$

One interesting observation about equation (28a) is that it is a completely
general description of the motion of particles or photons in a non-static
universe. As such, it also describes motion in a different model where we have
a
concentration of matter (rather than a hole) at the origin in an otherwise
uniform universe. Physical considerations, as well as superposition of the
regular Schwarzschild solution for an attracting mass onto the flat geometry of
section 4.2, indicate that matter is accelerated inward and the trajectories of
photons are curved inward toward the excess concentration of matter. In that
model, $C < 0, \l ' + \nu ' > 0,$ and $\dot\l (dr/ds)(dt/ds) < 0$. The
conclusion for that model is that $d^2r/ds^2 < 0$ for particles and photons. In
our model, $C > 0, \l ' + \nu ' < 0,$ and $\dot\l (dr/ds)(dt/ds) > 0$, and we
expect on exactly the same physical grounds that $d^2r/ds^2 > 0.$ That is,
particles and photons are repelled away from the hole, or are attracted toward
the mass in the spherical region S (see Fig. 4; note that coordinates in the
figure are spherical) which is symmetrically-placed on the opposite side of the
particle or photon from the hole.

\eject
\vbox{\vspace{6in}}
\special{Figs3.eps}

First, consider motion of a particle. Assume that it starts from rest ($dr/ds =
0$) and travels radially ($h = 0$). Equation (28a) can then be written as

$${{d^2r}\o{ds^2}} = -\dot\l \le({{dr}\o{ds}}\ri)\le({{dt}\o{ds}}\ri) -
{{e^{-\l}}\o 2}\le[\l 'e^{\l}\le({{dr}\o{ds}}\ri)^2 + \nu
'e^{\nu}c^2\le({{dt}\o{ds}}\ri)^2\ri]. \eqno (29)$$

\n By previous assumptions that $\l ' < 0, \nu ' < 0,$ and $dr/ds(0) = 0$, this
represents a positive initial acceleration. Therefore, $dr/ds$ initially
increases from zero. We conclude that $dr/ds$ and $d^2r/ds^2$ are both positive
on some initial interval $0 < t < t_1$. That is, particles are initially
accelerated outward away from the hole. Physical considerations indicate that
this situation will continue forever. On the other hand, if $d^2r/ds^2$ ever
turned negative, then $dr/ds$ would start to decrease, and the troublesome term
(the first one on the right) in equation (29) would decrease in significance.
In
fact, $dr/ds$ could never become zero with $d^2r/ds^2 < 0$, since that would
contradict equation (29), so we conclude that $dr/ds$ is always positive. We
infer that $d^2r/ds^2$ is also always positive.

Now consider the trajectory of a photon, with $h \neq  0$ but $ds = 0$. Let
$\mu$ be a parameter to replace $s$, increasing along the geodesic in the
direction of motion. The geodesic equation (28a) has the same form as that for
the motion of a particle, with the presence of the $h^2/r^2$ term. At the point
where $dr/d\mu  = 0$ (the point of closest approach to the origin), $d^2r/d\mu
^2 > 0$ by equation (28a). By the same reasoning as before, we deduce that
$dr/d\mu  > 0$ from that time forward and infer that $d^2r/d\mu ^2 > 0$ for all
time.

Perhaps it is not possible to actually prove mathematically that $d^2r/ds^2$ is
positive for particles and photons without a more detailed description of the
metric functions $\l$ and $\nu$. The main conclusion we wish to draw is that
particles and photons are accelerated outward, away from the hole. This result
is independent of the actual structure of the metric (within the restrictions
placed on it above). It only depends on $\l ' + \nu ' < 0$ which, in turn,
depends on $C > 0$, a result derived in section 4.3 when dealing with the
potential of a hole. We thus see that holes and gravitating bodies have exactly
the opposite effect on matter and energy.

\subsection{Summary of the theory}
\i We have shown, on the basis of the Newtonian theory, that a spherical hole
in
an otherwise infinite and uniform universe tends to repel matter. Using
Einstein's theory of general relativity above, we have come to the same
conclusion for both matter and photons, the quantum packets of radiation. Along
the way, we have developed some interesting formul{\ae} and concepts and have
considered Einstein's equations for a nonstatic line element and the
corresponding geodesics. In addition, we have given a physical explanation of
the cosmological constant and have shown the possibility of an infinite, but
unstable, universe in the general theory of relativity.

The consequences of these new discoveries in the field of cosmology are
staggering. Below, we develop an entirely new cosmology based upon holes and
their repulsive characteristics in an infinite universe. The mathematical
results given here form the basis for that development.

\section{Cosmology}

\i Now, we consider the implications of this theory to cosmology. We assume
that
our universe started out in an infinite, uniform configuration
infinitely long ago. Then we develop some of the logical consequences of such
an assumption and show that the resulting universe is surprisingly like our own
universe today. In particular, the expansion of the universe can be explained
in
terms of an accelerating expansion due to negative gravity rather that an
explosive but decelerating expansion due to the Big Bang. Of course, any
cosmology, especially one which reportedly refutes the Big Bang theory, must be
able to explain several astronomical observations such as the microwave
background radiation, the Hubble expansion, and the relative abundance of
hydrogen, helium, and other elements in the universe, among others. This is
done
also.

In section 5.1, the history of the universe is developed, starting with an
infinite, uniform state, and it is shown how structure could evolve in the
universe to its present state. In section 5.2, the gravitational redshift and
some of its consequences for the evolution of the universe are discussed.
Section 5.3 gives a comparison between this new cosmology and the Big Bang.

\subsection{History and structure of the universe}

\i We showed mathematically, using Newtonian mechanics, that there exists an
outward gravitational force acting at each point in an infinite universe with a
spherical hole whose center is at the origin. At any point at a distance $r$
from the center of the hole, the outward force on a mass $m$ is given by
equation (1). Then, we came to a similar conclusion using Einstein's theory of
general relativity. Using this idea that an outward force would cause holes to
expand, R. N. Lewis \cite{rlewis} showed how structure could form in a universe
which started out in a uniform state. The idea is that two spherical holes in
an
otherwise infinite and uniform universe would migrate toward one another, they
would eventually touch and form a three-dimensional figure-eight, and they
would
eventually merge into one larger sphere. The mass in between the two holes
would
be pushed aside. If three holes were in close proximity, they too would
approach
each other and eventually merge, pushing aside the mass between them. Four
holes
in close proximity to each other, however, retain the possibility that they
trap
some of the mass in between them as they move toward one another, thus forming
a
larger concentration of matter which would then pull together under its own
positive gravitational attraction.

We assume the existence of many holes throughout the universe, without
specifying how they came into being. Perhaps there were always slight
imperfections in the uniformity which, as we earlier pointed out, would
gradually grow in size. Another possibility for the introduction of holes is
some quantum effect which would destroy the total uniformity of the universe.
Whatever the cause, we assume that it repeated itself throughout the universe
to
form enough holes so that the ensuing evolution described below could occur.

These holes, or any regions of lower than average mass density, would expand
outward, just as lumps, or regions of higher than average mass density, would
attract other mass toward itself. In other words, the infinite, uniform
universe
is unstable.

We can imagine many expanding holes existing throughout the universe. R. N.
Lewis \cite{rlewis} referred to this configuration as the Swiss-Cheese
Universe. Later, as two holes came together, a sheet of matter would form
between them, and the sheets of matter, looking at the entire conglomeration,
would look like a foamy mixture, named the Foam Universe. Still later, as the
holes continue to expand and encroach upon each other's territory, the matter
in the sheets would migrate toward the edge of the sheet, forming filaments
connecting matter concentrations. This network configuration is referred to as
the Network Universe. Even later still, the filaments would split and be drawn
gravitationally toward a concentration of matter. This arrangement is referred
to as the Caltrop Universe because the appearance of one of these
concentrations
of matter is similar to a medieval caltrop.

When astronomers peer outward toward other galaxies, they notice just this type
of structure. Galactic clusters seem to be connected to each other by filaments
of galaxies, and, in turn, they appear clustered, not randomly, but in sheets,
just the structure predicted above. Galaxies and clusters are, in turn,
separated from each other by gigantic voids which occur naturally in the
above-described theory. It therefore appears that this view of the universe is
better able to explain the observed structure of the universe than is the Big
Bang theory, whose proponents are still arguing about how the universe could
have differentiated into stars, galaxies, and galactic clusters in the
relatively short time since the Big Bang.

What is necessary for the above evolution to have occurred as stated? The only
requirements are an infinite universe, matter distributed somewhat uniformly
throughout it, and the existence of randomly-scattered holes. The matter that
was once distributed uniformly has been pushed or pulled together into
galaxies,
surrounded by emptyness, that we observe when looking out into the universe
today. Assuming that these galaxies are distributed somewhat uniformly
throughout our universe, as most cosmologists believe today, then the outward
force of negative gravity must also exist at this time, tending to push the
universe apart.

We can imagine that the universe we observe (perhaps 10 billion light years or
more in radius) has an average mass density slightly less that that at
infinity.
One possible mechanism for this to happen is the following. Assuming that there
were events that occurred to form each of the holes initially all over the
universe, it is conceivable that the effect of those events sufficiently far
away from us haven't been felt yet on earth, so that we on earth observe the
average mass density in our neighborhood to be less than that farther out. Our
neighborhood (radius unspecified) of the universe and the galaxies in it,
therefore, effectively reside in a large hole, with the force of negative
gravity pushing outward. Of course, under this scenario of locality, this same
comment would apply to any observer situated at any point of the universe.
Therefore, the same lower-than-average mass density and the same expansion must
be observed everywhere.

\subsection{Gravitational redshift and its consequences}

\i The outward force due to negative gravity at any point was given above in
equation (1). In addition to that outward force, there is an additional inward
force due to the positive gravitational force of all mass closer to the center
of the hole than the mass [equ.(2)]. The total force was given earlier as
equation (8a). We also gave the corresponding potential as equation (8b).

It is well known that light escaping from a gravitating body undergoes a shift
in frequency due to the gravitational field \cite[sec. 4.4]{abs}. The shift is
given by

$$z = \D\nu /\nu _0 = \D\phi /c^2, \eqno (30)$$

\n where $\nu _0$ is the frequency of the emitted wave, $\D\nu $ is its change
in frequency, and $\D\phi $ is the change in potential.

In calculating $\D\phi $, there is a contribution from the emitting galaxy and
one from the observing galaxy (our own), each dependent on the gravitational
mass of the respective galaxies. For a galaxy of roughly the same size as our
own, these contributions will cancel. On the other hand, a much more massive
galaxy may exhibit a relatively large gravitational redshift. There is also a
Doppler-type effect due to relative velocity, which is currently believed to be
the primary cause of observed frequency shift. Assuming that the visible
universe resides in an expanding hole (see section 5.1), there will also be a
contribution from the potential of the hole due to negative gravity. This
Newtonian potential [see equ. (8b)] is proportional to $r^2$ for large $r$, and
could overwhelm the contributions from the other gravitational effects. It
could
do the same relative to Doppler effects as well.

It is apparent that whatever frequency shift is observed, at least four
causes must be taken into account in the analysis: frequency shift caused by
(1) gravitational effect of emitting galaxy, (2) gravitational effect of
absorbing galaxy, (3) gravitational effect of negative gravity, and (4) Doppler
effect due to relative motion. However, we can conclude that the effects of (3)
and (4) above reinforce each other, so that the observed redshift is not
entirely due to relative motion. In other words, the estimated velocity of
recession from observations of redshifted light is exaggerated, especially for
galaxies which are relatively far away. In the past, an observed redshift value
has been interpreted as indicating a certain distance and velocity of recession
(Hubble's Law). However, if (negative) gravitational effects as well as
relative
motion are taken into account, the observed redshift would indicate a slower
(less Doppler redshift) galaxy which is closer to us. The observed redshift
should not be attributed entirely to Doppler effects, nor entirely to
gravitational effects. The negative gravitational redshift becomes increasingly
important the larger that $r$ becomes. We note that, for $r = 10^9$ light
years,
equation (30) yields a gravitational redshift given by $2$ x $10^{-4}$, which
is
significant when dealing with galaxies at that distance. In observing distant
galaxies, this must be taken into account. For example, the Great Wall
\cite{gandh}, distances to which are predicted on the basis of observed
redshift
values, is at a distance where negative gravitational effects are significant.
More importantly, however, these negative gravitational effects will affect the
microwave background radiation (see below).

Because of the above comments, the entire Big Bang theory needs to be
reevaluated in light of negative gravity. An entirely new cosmology based upon
expansion due to negative gravity has been developed. R. N. Lewis \cite{rlewis}
shows one approach to this. In fact, his explanation gives a simple account for
the present structure of the universe which the Big Bang theory cannot explain
adequately without internal contradictions. A new cosmology was developed in
which large velocities (see section 3.2) are achieved due entirely to the
effects of negative gravity over much larger time intervals (many orders of
magnitude) than are currently believed to be required for expansion to our
present state. We also showed that Hubble's Law holds in this case. For this
cosmology, all that is required is that the visible universe, as well as
possibly much more, resides in a region of slightly smaller mass density than
that at infinity. This would allow negative gravity, as well as all of its
consequences, to function.

The observed abundances of helium, deuterium, and other heavier elements in the
universe can be explained by supernova activity during the longer time
intervals. It is currently believed that supernova activity since the Big Bang
was not sufficient to produce anywhere near the amount of heavier elements that
are observed in the universe. The amount of helium present is estimated to be
from $15$ to $40\%$  of the mass of the universe, with a preference for a
figure of $25\%$. It is thought that perhaps one-tenth of that amount could
have been produced in supernovas in the $15$ billion years since the Big Bang.
In a universe which has been around forever and which may have been nuclearly
active for $150+$ billion years, the observed abundances of helium and other
heavier elements could easily have been produced.

The uniform microwave background radiation could be accounted for by the
radiation of all galaxies. Our sun radiates at close to blackbody levels. We
expect the same of other stars and galaxies. Averaged out over regions of the
sky, we expect the radiation coming from other galaxies to be uniform and of
blackbody type. Indeed, data from the Cosmic Background Explorer \cite{mat}
support this conclusion.

It is generally agreed that the main nonuniformity in the background radiation
is due to the fact that we are in motion relative to the source of that
radiation. Most cosmologists believe that the motion is due to the Big Bang.
The
motion could be due to a general expansion caused by negative gravity.

We also expect the radiation from other galaxies to be redshifted because of
relative motion. From the discussion above, it is apparent that radiation is
also redshifted because of the negative gravitational potential. This
redshifted
radiation is observed in the microwave region. It is interesting to note that
the singularity implicit in equation (22b) (which gives the metric tensor:
$e^{\nu} = e^{-\lambda} = 1 - \Lambda r^2/3$),

$$r_s = \sqrt{{3\o \Lambda}} = 3.11 \mb{x} 10^{26} \mb{m} = 3.29 \mb{x} 10^{10}
\mb{ly}, \eqno (31)$$

\n yields a redshift from equation (30) of

$$z = 0.25, \eqno (32)$$

\n this value being independent of $r_s, \Lambda,$ or $\rho$. If the microwave
background radiation is actually due to an integration of the radiation of all
galaxies, it would indicate a redshift of $z \approx  2000$, corresponding to a
temperature of 2.73 K and stellar surface temperatures on the order of 5000
K (possibly less in the cooler, earlier universe). Thus, it seems likely that
the origin of the microwave background radiation must be $10$ to $100$ times
more distant than $r_s$, if we are to attribute the redshift to gravitational,
rather than Doppler, effects. More likely, it is due to a combination of the
two
effects. The singularity needs to be investigated further. However, it must be
remembered that our universe is not exactly as modeled, so that the singularity
shown above for a uniform universe with a spherical hole may not be a
singularity for an uneven universe such as our own.

The value $z = 0.25$ found above is larger than most observed galactic
redshifts. However, some recent observations of quasar redshifts are on the
order of $3.5 - 4$. These can be explained by Doppler effects or positive
gravitational effects due to incredibly high concentrations of mass, possibly
in
combination with the negative gravitational redshift. Our sun's gravitational
redshift is about $2$ x $10^{-6}$. A redshift of $z = 4$ could result from a
star or quasar with mass-to-radius radio of $2$ x $10^6$ times that of our sun.
Weinberg \cite[equ. (11.6.20)]{wei} gives an upper bound of $z = 2$ for
gravitational redshift in a universe with zero cosmological constant. As can be
seen from the equation before Weinberg's equation (11.6.20), using our equation
(11) for his $B(r)$, a positive $\Lambda$ mediates a higher value of $z$.

Of course this view of cosmology denies the Great Cosmological Principle of
isotropy and homogeneity and the principle of covariance by giving preference
to centers of holes, at least at the present time. If we travel backward in
time, the universe as has been presented here should appear more homogeneous
and colder, so that these principles would have been more valid at earlier
times. But, of course, as we look outward toward the heavens, we see that the
universe is not isotropic nor homogeneous, so this should not be of concern.

Olbers' paradox has an interesting explanation in this new theory of cosmology.
Wesson \cite{wes} gives an explanation based upon the finite age of the
universe and on inflation of the universe. This new theory does not require
inflation or the Big Bang, but we can still invoke Wesson's argument of the
finite age of the universe by stating that, at earlier times, the universe was
less radiant (less hot). Furthermore, another interesting explanation lies in
the interpretation of the light ray geodesics. These bend outward, so that
radiation from far enough away never reaches us!

One final comment deals with whether the universe is open or closed. This and
the structure of the universe seem to be two of the most important open
questions in the study of cosmology. In any cosmology based upon negative
gravity, the universe is not only open, but its expansion is accelerating
outward. There is not any great mass pulling it inward. In fact, the ``mass at
infinity" is pulling it apart.

\subsection{Comparison with the Big Bang}

\i Why should we believe in a cosmology based upon negative gravity rather than
one based upon the Big Bang? The Big Bang theory was developed to explain the
observed redshift of light coming from distant galaxies. This redshift, in
turn, was explained by attributing it to recessional motion, mainly because no
other explanation for the observed redshift was known. This motion was then
attributed to a general expansion of the universe. It was argued that this
expansion could only be explained by extrapolating back to a time when an
explosion (Big Bang) caused the motion we apparently observe today. No other
mechanism for expansion was known. The subsequent discovery of the microwave
background radiation fit in with this theory very nicely. This radiation was
thought to be the remnant radiation left over from the Big Bang. It was further
deduced that the expansion must be decelerating due to the inward pull of the
finite mass distribution. The theory was further refined to include other
recent observations, including the presence and concentration of heavy
elements. The theory still has trouble explaining the observed structure of the
universe, including the presence of large voids and the clustering of galaxies.
Recent observations from the Cosmic Background Explorer \cite{mat} show the
``early universe" to be almost perfectly uniform and homogeneous, not leaving
enough time for matter to have differentiated and formed into the universe that
we observe today.

The theory of the development of the universe presented here, which is based on
negative gravity in an infinite universe, does not have this shortcoming. It
predicts the gross structure of the universe on the basis of mass being pushed
or pulled together by the combined negative gravitational effects of adjacent
holes \cite{rlewis}. In addition, it presents a viable alternate explanation of
the observed redshift of galactic radiation and of the microwave background
radiation. The observed redshift of galactic light can be explained as due
partly to recessional motion of expansion and partly to gravitational redshift,
both (expansion and gravitational redshift) caused by the negative
gravitational potential of ``mass at infinity". It is possible to separate out
the contribution to the redshift $z$ due to negative gravity
and that due to recessional motion, assuming that some form of Hubble's Law is
still valid. In fact, earlier, we showed that $v = \beta r$
(velocity proportional to distance) for such a universe, which shows that
Hubble's Law (with constant $\beta$) is valid, without the necessity of
extrapolating back in time to a singularity. In addition, the microwave
background radiation can also be associated with a redshift of radiation due
to the negative gravitational potential. Finally, the abundances of elements
other than hydrogen can be explained by supernova activity over the incredibly
long life of the luminous universe (at least an order of magnitude larger than
is currently believed). This new theory predicts an expansion which originated
rather benignly in a cold, (relatively) motionless, infinite universe and which
has accelerated outward and will continue to accelerate outward forever.

\section{Calculation of values of the Hubble constant, the average mass density
of the universe, and the cosmological constant}

The purpose of this section is to use some of the
equations derived in previous sections to deduce accurate values for the
average mass density of the universe, $\rho _0$, Hubble's constant, $H_0$, and
the cosmological constant, $\L$. These calculations are based upon fairly
accurate data, including redshift and distance data to specific galaxies. The
computations can be done for any such galaxies, but we concentrate on one,
M87, for a variety of reasons. Section 6.1 details these reasons and the
specific data we use. Section 6.2 includes the calculations of the different
constants. Finally, in section 6.3, a summary and conclusions as well
as justification of the methods developed here are presented.

\subsection{Why study M87?}

\i In order to perform the calculations of the values of the particular
constants, specific astronomical data are needed. There are several very good
reasons why the galaxy M87 would yield reliable data on which the conclusions
we reach are based.

First, the redshift of radiation from M87 has been reasonable accurately
measured. The value for that redshift is about $z = 0.0041$ \cite[p. 639]{wei}.

Second, it is close enough to us that its distance from us can be fairly
accurately measured by standard (not redshift) means. This measured distance is
$55-65$ million light years \cite[p. 439]{wei}. We will use both figures in our
calculations. Any error in measuring this distance will show up later in our
calculation of the values of the constants. M87 is also close enough that small
$z$ approximations should be valid.

Third, it lies near the center of the Virgo Cluster of galaxies, and is
relatively motionless with respect to that cluster. Therefore, such motions
caused by local gravitational effects (such as rotation about the center) are
not an added complication. We assume that the entire redshift, $z$, is due to
universal expansion (see above for more details; this expansion is not
due to the Big Bang), and that local gravitationally-induced motions are
insignificant. M87 is far enough away that this seems to be a reasonable
assumption.

Fourth, M87 lies in the general direction toward which our galaxy and galactic
cluster are moving. The observation of our direction of motion is based upon
relative microwave background radiation measurements. For this reason, the
calculations based upon negative gravity should be relatively accurate on the
assumption that the Virgo Cluster is farther out than we are along a radial
direction from the center of expansion.

One drawback of using M87 is that it is a huge galaxy relative to our own,
roughly $500$ times as massive \cite[p. 95]{rub}. As such, it may have
associated with it a significant gravitational redshift compared with ours,
which may have to be accounted for in the calculations. Ignoring this possible
gravitational redshift, as we do here, means that the results for $z_m$ and
$z_n$ later in this paper are upper bounds and may need to be reduced, thus
changing the other calculated values.

One problem associated with the use of data from only one galaxy is that the
galaxy we choose may not be typical. A more comprehensive study could be done
using a statistical analysis of data from many different galaxies, with proper
precautions being taken to account for, in particular, the phenomenon referred
to in the fourth category above. However, the fact that our calculation below
gives a value of the Hubble constant in the middle of the
experimentally-determined range should convince the reader that M87 is a
reasonable choice and that this approach is a sound one.

\subsection{Calculation of $\rho _0$, $H_0$, and $\L$}

\i First, using the notation $z$ for redshift, we identify three causes of
redshift: (1) redshift due to recessional motion, $z_m$, (2) redshift due to
negative gravity, $z_n$, and (3) gravitational redshift, $z_g$. For a given
value of $z$, the wavelength of the radiation has been shifted by a factor of
$1
+ z$. Therefore, $1 + z = (1 + z_m)(1 + z_n)(1 + z_g)$. We assume that $z_g$ is
negligible, so that the above equation can be replaced by $1 + z = (1 + z_m)(1
+
z_n)$ or

$$z = z_m + z_n + z_mz_n. \eqno (33)$$

\n But,

$$cz_m = v = rH_0, \eqno (34)$$

\n where $H_0 = \be$ represents Hubble's constant, and its representation as
found above equation (4) is

$$H_0 = \sqrt{{{4\pi G\rho _0}\o 3}}. \eqno (35)$$

\n In these equations, $c$ is the speed of light in a vacuum, $v$ is the
velocity of recession, $r$ is distance, $\rho _0$ is mass density, and $G$ is
the gravitational constant. Therefore,

$$z_m = {v\o c} ={{rH_0}\o c} = \sqrt{{{4\pi G\rho _0r^2}\o{3c^2}}}. \eqno
(36)$$

\n Using the assumptions given above,
particularly equations (30, 34, 35), we find

$$z_n = {{4\pi G\rho _0r^2}\o{6c^2}} = {{z_m^2}\o 2}. \eqno (37)$$

\n Therefore, from equation (33),

$$z = z_m + {{z_m^2}\o 2} + {{z_m^3}\o 2}. \eqno (38)$$

\n For the galaxy M87, $z = 0.0041$, so that equations (38) and (37) imply

$$\begin{array}{c}
z_m = 0.0040916,\\
\\
z_n = 0.0000084.

\end{array} \eqno (39)$$

\n Using $r = 6.5$ x $10^7$ l.y., equation (37) yields,

$$\rho _0 = {{6c^2z_n}\o{4\pi Gr^2}} = 1.43 \ \mb{x} \ 10^{-26} \mb{kg/m}^3.
\eqno (40a)$$

\n Using $r = 5.5$ x $10^7$ l.y., we find

$$\rho _0 = 2.00 \ \mb{x} \ 10^{-26} \mb{kg/m}^3. \eqno (40b)$$

Since the $r$ values are line-of-sight distances and are probably greater than
the difference in distance from the center of the main expansion, the $r$
values should be reduced by an indeterminant amount, thus resulting in a larger
value for $\rho _0$. The approximate value of $\rho _0$ used earlier was $1.67$
x $10^{-27}$ kg/m$^3$, roughly one-tenth the value
calculated above. Rubin \cite{rub} finds that the average value of $\rho _0$,
at least in our visible universe, is $6.73$ x $10^{-27}$ kg/m$^3$. This figure
is based upon gravitational calculations and is about $1000$ times larger than
observations of luminous matter lead us to believe. Thus, Rubin's calculations
indicate that a large percentage of the matter in the universe is nonluminous.
If these are accurate figures, they indicate that the average mass density at
infinity is roughly $2-3$ times what it is in our neighborhood, which is in
line
with the assumption made above and referred to in section 5.

Using the range of values for $\rho _0$ given in equations (40), equation (35)
yields the range of values for Hubble's constant:

$$H_0 = \sqrt{{{4\pi G\rho _0}\o 3}} = (2.00-2.36) \ \mb{x} \ 10^{-18}
\mb{s}^{-1}. \eqno (41)$$

\n We note that the range of values we derive by this method for Hubble's
constant is right in the middle of the range of values currently used for that
value. It corresponds to the value $62-73$ km/s/Mpc. This correlation of values
must lend credence to this line of reasoning. It is extremely unlikely that we
would end up with this degree of accuracy, compared with
experimentally-determined values, if the underlying theory were not correct.

Finally, the range of values of the cosmological constant is

$$\L  = {{4\pi G\rho _0}\o{c^2}} = (1.33-1.86) \ \mb{x} \ 10^{-52} \mb{m}^{-2}.
\eqno (42)$$

\n Although this value is extremely small on the scale of our solar system or
even our galaxy, it nevertheless is increasingly important on larger and larger
scales.

\subsection{Summary and conclusions}

\i In this section, we have used certain astronomical data, certain
well-established formul{\ae}, and several new formul{\ae} for calculating the
values of the average mass density of the universe, $\rho _0$, the Hubble
constant, $H_0$, and   the cosmological constant, $\L$. These values are found
to be in the ranges:

$$\rho _0 = (1.43-2.00) \ \mb{x} \ 10^{-26} \mb{kg/m}^3,$$

$$H_0 = (62-73) \mb{km/s/Mpc}, $$

$$\L  = (1.33-1.86) \ \mb{x} \ 10^{-52} \mb{m}^{-2}.$$

The fact that $H_0$ lies exactly in the middle of the experimentally-determined
range of values for that constant must mean several things. First and most
importantly, it signifies that the underlying theory of negative gravity must
be correct. Otherwise, it would be exceedingly improbable that the calculated
value would be anywhere near the experimental value. In fact, we use formula
(37) above which is based upon a gravitational potential which is on the order
of $r^3$ times larger than the one normally used in astronomical calculations.
With the values of $r$ used here, the difference is about $70$ orders of
magnitude. If the underlying theory were not correct, the calculated values
should be different to a corresponding degree. Second, it shows that our use of
the galaxy M87 was a reasonable choice. Third, it serves to pin down the value
of the Hubble constant even closer. Fourth, the values given for the
cosmological constant and for the average mass density of the universe should
be
relatively accurate. It should be pointed out that this is the first attempt to
actually assign a nonzero value to the cosmological constant and to calculate
the average mass density in an infinite universe.

Finally, let us observe that, in the above calculations, there were several
places where it was indicated that certain values might need to be reduced for
various reasons, while, in other places, just the opposite was indicated.
Again,
since the calculated value of the Hubble constant falls in the
experimentally-determined range, it is reasonable to conclude that the above
mentioned changes approximately balance each other.

\bibliographystyle{plain}

\end{document}